\documentclass{article}
\usepackage{spconf,amsmath,graphicx}

\usepackage{etoolbox,siunitx}
\usepackage{multirow}
\usepackage{etoolbox,siunitx}
\robustify\bfseries
\usepackage{booktabs}
\usepackage{balance}
\usepackage{amssymb}
\usepackage{bm}
\usepackage{arydshln}
\usepackage{caption} 
\usepackage{xcolor}
\usepackage{enumitem, setspace, hyperref}
\usepackage{subcaption}
\usepackage{cleveref}

\DeclareMathAlphabet{\pazocal}{OMS}{zplm}{m}{n}
\newcommand{\Loss}{\pazocal{L}}

\title{HYPERBOLIC DISTANCE-BASED SPEECH SEPARATION}
\name{Darius Petermann$^1$ and Minje Kim$^2$\thanks{
This material is based upon work supported by the National Science Foundation under Grant No. 2046963.}
\sthanks{Work done at Indiana University.}}
\address{$^1$Indiana University, Department of Intelligent Systems Engineering, Bloomington, IN, USA 47408\\
$^2$University of Illinois at Urbana-Champaign, Department of Computer Science, IL, USA 61801}

\begin{document}
\ninept
\maketitle
\setlength{\abovedisplayskip}{2.5pt}
\setlength{\belowdisplayskip}{2.5pt}
\begin{abstract}

In this work, we explore the task of hierarchical distance-based speech separation defined on a hyperbolic manifold. Based on the recent advent of audio-related tasks performed in non-Euclidean spaces, we propose to make use of the Poincar\'e ball to effectively unveil the inherent hierarchical structure found in complex speaker mixtures. We design two sets of experiments in which the distance-based parent sound classes, namely ``near'' and ``far'', can contain up to two or three speakers (i.e., children) each. We show that our hyperbolic approach is suitable for unveiling hierarchical structure from the problem definition, resulting in improved child-level separation. We further show that a clear correlation emerges between the notion of hyperbolic certainty (i.e., the distance to the ball's origin) and acoustic semantics such as speaker density, inter-source location, and microphone-to-speaker distance.

\end{abstract}

\section{Introduction}
\label{sec:intro}

Humans can focus on a sound source of interest in complex acoustic scenes with remarkable ease \cite{mcdermott2009cpp}. Such ``selective attention" \cite{Karns2015AuditoryAI} could be based on contextual and subjective motivations, but other more established ones can improve source separation systems. By using the deep neural network (DNN)-based supervised learning approaches \cite{WangDL2018ieeeacmaslp} as the framework, 
recent research has attempted to encompass the selective attention concept by conditioning the model with the auxiliary information, such as text queries \cite{liu2022separate} or language, gender, and spatial information about the source \cite{tzinis2022heterogeneous}. In this paper we focus on the spatial cues \cite{darwin2000selective}, e.g., speaker positions or directional information, which have been proven to be useful \cite{rongzhi2019spatial,heitkaemper2019position,EskimezSE2021personalized}. 

More recently, the task of distance-based source separation was introduced \cite{Patterson2022DistSS}, whose goal is to separate a monoaural mixture $x$ into its distance-based constitutive components, the sum of near sources  $x^{(\mathcal{N})}=\sum_{k\in \mathcal{N}} s^{(k)}$ and far sources $x^{(\mathcal{F})}=\sum_{k\in \mathcal{F}} s^{(k)}$, where $\mathcal{N}$ and $\mathcal{F}$ stand for the sets of near and far sources, respectively. The grouping is defined by a distance threshold $\tau$. In \cite{Patterson2022DistSS}, the input is restricted to single-channel signals while proposing the speaker's distances to the microphone as a determining factor in the source separation task, differently from the previous work, such as SpeakerBeam \cite{Delcroix2018speaker_beam}, where the spatial information mainly comes from a multi-channel context.

\begin{figure}[t]
    \centering
        \includegraphics[width=0.95\linewidth]{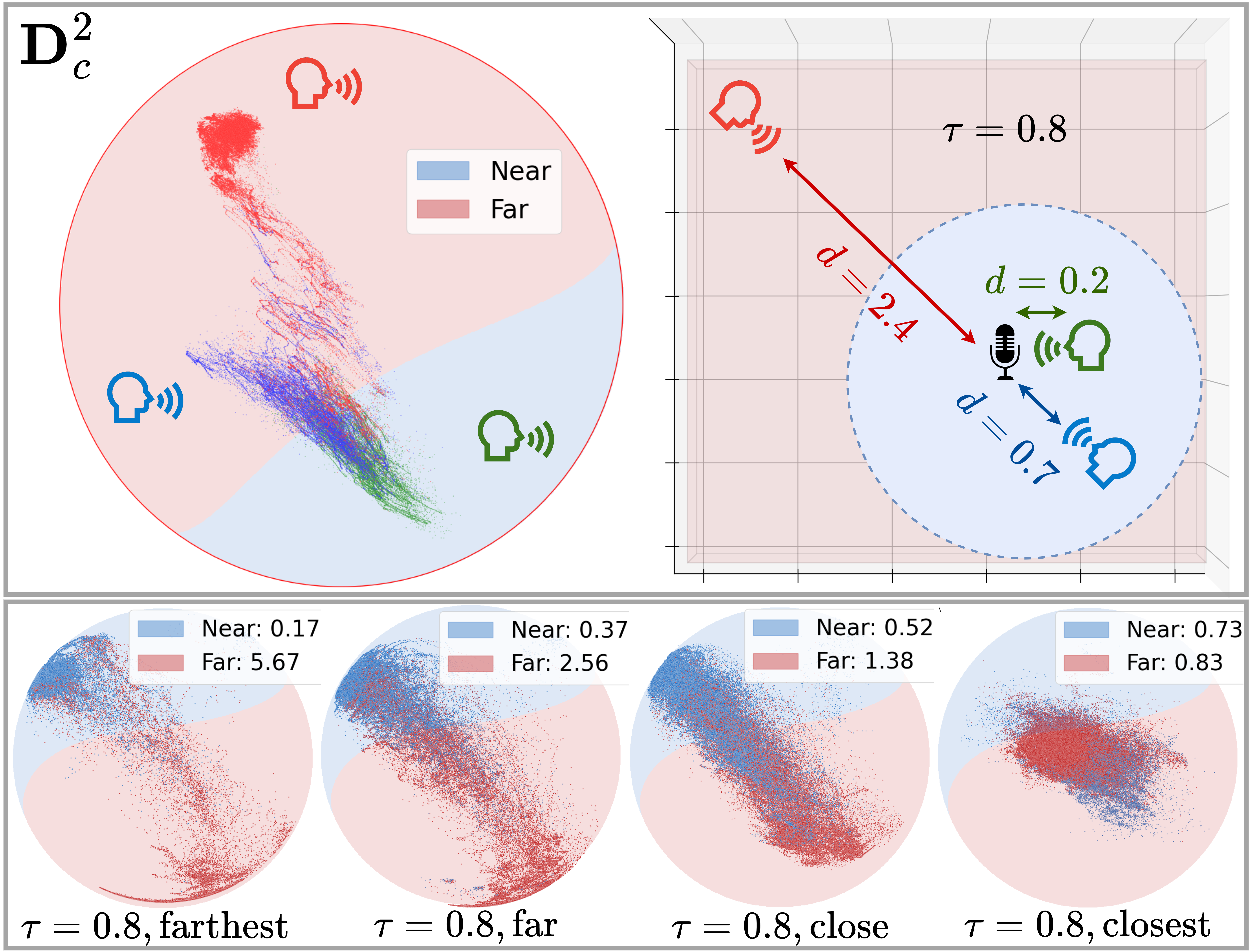}\vspace{-.1cm}
    \captionsetup{belowskip=-.5cm}
    \caption{Illustration of distance-based source separation performed on the Poincar\'e ball. Top-left is a projection of a three-speaker mixture, where two speakers (green and blue) belong to the near field ($\leq\!\tau=\!0.8$ meter) while one (red) to the far field ($>\!\tau$). Top-right shows the room configuration of the same mixture. Note that the source at the near-far field boundary tends to be projected at around the center of the Poincar\'e ball, reflecting the model's uncertainty. The bottom plots denote projections of two speakers that are progressively placed closer to $\tau$ (thus no hierarchy).}
    \label{fig:main}
\end{figure}

In this work, we propose to formulate a hierarchical version of the distance-based source separation problem. We present a two-level hierarchy, where the system separates the mixture into two parent nodes, $x^{(\mathcal{N})}$ whose source components are closer to the microphone than $\tau$, i.e., $||\bm{m}-\text{loc}(s^{(k)})||_2<\tau$, where $\bm{m}$ and $\text{loc}(s^{(k)})$ respectively denote the microphone and source locations in the 3D room, and vice versa for the farther source group $x^{(\mathcal{F})}$. In addition, the child-level task aims at separating individual sources $s^{(k)}$.

Inspired by the recent success of hyperbolic manifolds to model hierarchical structures in computer vision \cite{Atigh_2022_CVPR, Khrulkov_2020_CVPR}, natural language processing \cite{shimizu2021hyperbolic}, and more recently audio \cite{Petermann2023ICASSP_hyper,Nakashima2022VAE, GermainF2023hyperbolic}, we propose to utilize a popular hyperbolic model, the \emph{Poincar\'e ball} whose embeddings efficiently preserves the intrinsic hierarchical structure of the data in low dimensional space with minimal distortion \cite{sarkar2011distortion}.

The Poincar\'e ball has also proven to provide reliable and natural measures of certainty estimation \cite{Atigh_2022_CVPR,Khrulkov_2020_CVPR,suris2021hyperfuture}, which we can interpret as the model's source-specific confidence level put in the context of audio source separation \cite{Petermann2023ICASSP_hyper}. 
In our hierarchical setup, we assume that the geometry between the sources and the microphone introduces uncertainty to the parent-level separation task, i.e., the closer individual sources are to the threshold $\tau$, the more challenging the parent-level separation will be.
In addition, the child-level separation faces the challenges that a separation system typically needs to deal with, e.g., same-gender sources being more difficult to separate and therefore inducing additional uncertainty to the separation task. Yet, the child-level sources' geometry also contributes to the uncertainty mapping. Fig.~\ref{fig:main} exemplifies this paradigm; the close proximity of the blue source ($d\!=\!0.7$) to the threshold ($\tau\!=\!0.8$) leads its projections on the Poincar\'e ball to be closer to the origin, while the green source's proximity to the microphone ($d\!=\!0.2$) mitigates the uncertainty, thus pushing its projections to the edge of the ball. In this case the main cause of the performance degradation in the parent-level separation task is the blue source's location which puts many of its embeddings on the other side of the decision boundary. Meanwhile, the red source's farthest location pushes its embeddings away from the origin and consequently the near-far decision boundary as well, signifying the relevance of the geometry to both levels.

Our paper empirically demonstrates two conjectures. First, the hyperbolic model does a better job at hierarchically separating the parent and children sources in terms of scale-invariant signal-to-distortion ratio (SI-SDR) \cite{LeRouxJL2018sisdr} than its Euclidean counterpart. Second, we analytically show that the network can intrinsically map out the hyperbolic certainty on the Poincar\'e ball to a few acoustic semantics. We envision that the interpretability could benefit geometry-sensitive real-world applications, e.g., disambiguating meeting participants from other interferences based on their physical locations (i.e., inside the same meeting room or not), discerning close and important sound events from too-far-to-be-relevant ones, etc. We further claim that the simplicity of the projection-based conversion to the Poincar\'e ball makes it an efficient alternative to the sampling or ensemble-based uncertainty quantification method \cite{fang2023uncertainty}.

\section{HYPERBOLIC DISTANCE-BASED SPEECH SEPARATION}

\subsection{Problem Formulation}
\label{sec:problem_formulation}

Following \cite{Petermann2023ICASSP_hyper}, we formulate our source separation task as a mask-inference problem. Given a complex mixture spectrogram $X \in \mathbb{C}^{T\times F}$ with $K$ underlying sources and a separator model $f(\cdot)$, we estimate a $K$-dimensional mask $M \in \{0,1\}^{T\times F \times K}$, which is a one-hot vector and whose $k$-th element of a time-frequency (TF) bin $M_{t,f,k}=1$ only if the $k$-th source is the most dominant one in that bin. 
Likewise, the problem boils down to a $K$-class classification, where
softmax applied to $f(\cdot)$ infers the posterior probability of each TF bin $(t,f)$ belonging to the sources, i.e., $\hat{M}_{t,f,k}=P(k|X_{t,f})$. The mask estimates $\hat{M}$ are then element-wise multiplied with the original mixture $X$ in order to retrieve the $k$-th spectrogram estimate. The inverse short-time Fourier transform (iSTFT) converts it back to the time domain: $\hat{s}^{(k)} = \text{iSTFT}(X \odot \hat{M}_{:,:,k})$.
An established method to obtain these masks is to compute embeddings $Z_{t,f,:} \in \mathbb{R}^L$ of certain dimension $L$ for each of the TF bins, using a neural network-based transformation function, i.e., $Z\leftarrow f(X)$.

In the proposed system, the separator network projects the input to a hyperbolic manifold, i.e., the Poincar\'e ball, to perform classification on it \cite{Petermann2023ICASSP_hyper}. To this end, $f(\cdot)$ is repurposed: it originally consists of a bidirectional long short-term memory (BLSTM) stack followed by a linear layer that converts the BLSTM output to $Z$. As for the hyperbolic counterpart, $Z$ goes through an additional deterministic projection function $H=\exp^c_0(Z_{t,f,:})$ as in Eq.~\eqref{equation:proj}, followed by the hyperbolic multinomial logistic regression (HMLR) to substitute softmax. 
The whole process can be summarized as follows:
\begin{equation}
    P(k|X_{t,f})\approx \left\{\begin{array}{rl}
    \text{softmax}(Z_{t,f,:})      &  \text{Euclidean},\\
    \text{HMLR}(H_{t,f,:})      & \text{Hyperbolic.}
    \end{array}\right.
\end{equation}

Given two parent groups $x^{(\mathcal{N})}=\sum_{k\in\mathcal{N}}s^{(k)}$ and $x^{(\mathcal{F})}=\sum_{k\in\mathcal{F}}s^{(k)}$ as well as their respective children $s^{(k\in\mathcal{N})}$ and $s^{(k\in\mathcal{F})}$, source separation aims to estimate two sets of masks for parent and children levels, separately. The parent-level masks are $\hat{M}^{\text{parent}}_{t,f,0}\!=\!P(x^{(\mathcal{N})}|X_{t,f})$ and $\hat{M}^{\text{parent}}_{t,f,1}\!=\!P(x^{(\mathcal{F})}|X_{t,f})$, thus $\hat{M}^{\text{parent}}_{t,f,0}\!+\!\hat{M}^{\text{parent}}_{t,f,1}\!\!\!=\!\!\!1$. For the child level, an individual separation task per parent, i.e., $\sum_{k\in\mathcal{N}}\hat{M}^\text{near}_{t,f,k}\!\!=\!\!1$ and 
$\sum_{k\in\mathcal{F}}\hat{M}^\text{far}_{t,f,k}\!=\!1$, respectively, is defined. We assume independence between the two hierarchical levels, so the \emph{hierarchical softmax} produces level-specific mask estimates for the parents and children independently. 
\subsection{Hierarchy in the Hyperbolic Space}
Hyperbolic spaces have the inherent ability to efficiently represent tree-like structures in a continuous way \cite{nickel2017poincare}. Because of the unique geometry induced by these spaces, the distance from the origin grows exponentially (i.e., as a function of the curvature) with the radius, reaching infinity at the ball's edge.

Let $\mathbb{D}^n_c$ be the $n$-dimensional hyperbolic space with a Riemannian geometry and negative curvature $c<0$, which defines a cone-shaped hyperbolic manifold, while the other two isotropic model spaces being Euclidean ($c=0$) and spherical ($c>0$). In this paper, we use the Poincar\'e ball, one of the most popular hyperbolic models for gradient-based learning \cite{Atigh_2022_CVPR,suris2021hyperfuture,Khrulkov_2020_CVPR,Petermann2023ICASSP_hyper}, defined by the manifold and Riemannian metric pair $\mathcal{M}$ and $g^\mathbb{D}_c(x) = (\lambda^c_x)^2g^E$, respectively, where $\lambda^c_x=2/(1-c\|x\|^2)$ is a so-called \emph{conformal factor} and $g^E$ the Euclidean metric. The model defines its \emph{induced distance} metric $d$ in relation to the center of the ball: 
$d_{\mathbb{D}}(a,b)  = \operatorname{cosh}^{-1} \left( 1 + 2\frac{||a-b||^2}{(1-||a||^2)(1-||b||^2)} \right)$,
where $d_{\mathbb{D}}(a,b)$ defines the shortest distance path, or \emph{geodesic}, between points $a$ and $b$ on the Poincar\'e ball. In the context of the hierarchical trees, the mean between two leaves in this space would result in a parent rather than another intermediate leaf, and would be located closer to the ball's origin (i.e., analogous to a continuous hierarchy).
\subsection{Hyperbolic Learning}

Our architecture operates in both Euclidean and hyperbolic spaces through exponential and logarithmic maps, allowing us to project the embeddings to and from the hyperbolic space, respectively. For embeddings $v \in \mathbb{R}^n \setminus \{0\}$ and $y \in \mathbb{D}^n_c \setminus \{0\}$,
\begin{equation}
    \exp^c_0(v) = \frac{\text{tanh}(\sqrt{c}\|v\|)}{\sqrt{c}\|v\|}v, \quad \log^c_0(y) = \frac{\text{tanh}^{-1}(\sqrt{c}\|y\|)}{\sqrt{c}\|y\|}y,
    \label{equation:proj}
\end{equation}
where the former represents the deterministic conversion applied to the output of the separator network, i.e., $H_{t,f}=\exp^c_0(Z_{t,f})$, as a mapping from the tangent space to the hyperbolic manifold, while the latter the inverse (i.e., $\mathbb{R}^n$ to $\mathbb{D}^n_c$ and inversely). Note that both projections are centered around their subscript 0. Once the embeddings are projected into the hyperbolic space, the key operations such as vector additions and multiplication follow the rules derived by the Riemannian metric in the Poincar\'e space \cite{shimizu2021hyperbolic}.

In Euclidean geometry, MLR works with the assumption that logits are the representation of the embedding's distance to the class hyperplanes. In the hyperbolic space, this assumption can be reformulated from the perspective of margin hyperplanes. For a given class $k \in \{1,\ldots,K\}$, one can define a geodesic in $\mathbb{D}^n_c$ orthogonal to $a_k$ and containing $p_k$. Considering an embedding $Z_{t,f}$ projected onto the ball, the final hyperbolic MLR formula is:
    $P(k|H_{t,f}\!)\!\propto\!\exp\Bigl(\!\frac{\lambda^c_{p_k}\|a_k\|}{\sqrt{c}} \text{sinh}^{\!\!-1} \!
    \Bigl( \frac{2\sqrt{c}|\langle -p_k\!\oplus_c\! H_{t,f},a_k \rangle|}{(1\!-\!c\|\!-\!p_k \!\oplus_c\! H_{t,f})
\|^2)\|a_k\|} \Bigr)\!\!\Bigr)$,
where $\oplus_c$ denotes the M\"obius addition in $\mathbb{D}^n_c$ \cite{chen2022fully}. Note that both $p_k$ and $a_k$ are learnable parameters.

\section{Data Curation and Simulation}

To validate our method with various acoustics configurations, we use similar methods as presented in \cite{Patterson2022DistSS}. 
We first generate random room impulse responses (RIRs) using the Pyroomacoustics toolbox \cite{scheibler2018pyroomacoustics}. The room dimensions are uniformly randomized between $3.0 \times 4.0 \times 2.13$ and $7.0 \times 8.0 \times 3.03$ meters. Microphone locations are randomly sampled within the resulting room dimension. In each room, four source locations are independently and randomly sampled from a beta distribution based on their distance from the microphone.
The target $RT_{60}$ (in seconds) for each room simulation is randomly sampled in a similar fashion from the range $[0.1, 0.5]$. For the speech sources, we opt to use Libri-light \cite{librilight} to generate the training set, and Librispeech-clean \cite{PanayotovV2015Librispeech} for validation and testing. 

During training, a room is first randomly sampled without replacement for each training example. Six-second-long speech chunks are then randomly sampled from the pool of speech utterances with replacement by zero-padding short utterances. For each training mixture, all four sources are expected to be exploited unless their associated parent class has already reached the maximum number of allowed children. 
The validation and testing sets are prepared using the same procedure.

\begin{table}
    \scriptsize
    \centering
    \sisetup{table-format=2,round-mode=places,round-precision=0,table-number-alignment = center,detect-weight=true}
    \setlength\tabcolsep{4.0pt}
    \captionsetup[table]{skip=5pt}
    \begin{tabular}{lSSSSSSS}
    \toprule
    \textbf{Density ($2$ children)} & \{{2,0}\} & \{{2,1}\}  & \{{2,2}\}  & \{{1,2}\}  & \{{0,2}\} \\
    \textbf{Density ($3$ children)} & \{{3,0}\} & \{{3,1}\}  & \{{2,2}\}  & \{{1,3}\}  & \{{0,3}\} \\
    \midrule
    \textbf{Training} & \num{3156} & \num{8217} & \num{28059} & \num{71152} & \num{89416} \\
    \textbf{Validation} & \num{45} & \num{101} & \num{321} & \num{916} & \num{1117} \\
    \textbf{Testing} & \num{400} & \num{400} & \num{400} & \num{400} & \num{400} \\
    \bottomrule
    \end{tabular}
    \captionsetup{belowskip=-.6cm}
    \caption{Distribution of the various speaker density configurations across our two hierarchical use-cases.}
    \label{table:dataset_stats}
\end{table}

\section{Experimental Settings}

To validate our approach, we design a set of three experiments; the first one solely aims at separating the parent sources $x^{(\mathcal{N})}$ and $x^{(\mathcal{F})}$ (i.e., one-level model) as in \cite{Patterson2022DistSS}. The latter two exploit the two-level models with different hierarchical configurations: one assumes a maximum of two child sources per parent, while the other allows three child sources per parent. Table~\ref{table:dataset_stats} denotes the five possible speaker setups found in our dataset for each configuration.

\noindent\textbf{Model architecture}:
Our model follows the same architecture as described in \cite{Petermann2023ICASSP_hyper}, which consists of four BLSTM layers with 600 units in each direction; a dense layer is then used to obtain an Euclidean embedding for each TF bin where $L=2$. A dropout of rate $0.3$ is applied to the output of each BLSTM layer, except the last one. 
For the hyperbolic model with negative curvature ($c\!<\!0)$, an additional exponential projection layer maps the Euclidean embeddings onto the Poincar\'e ball. As discussed in Section~\ref{sec:problem_formulation}, either softmax with Euclidean logits or the hyperbolic MLR version then computes the masks for each of the source classes. In practice, we follow the hierarchical softmax approach from~\cite{Atigh_2022_CVPR}, and therefore have two MLR layers: one for the binary classification of the parent sources $x^{(\mathcal{N})}$ and $x^{(\mathcal{F})}$, and the second one with $K\!=\!4$ or $K\!=\!6$ for the leaf classes, as dictated by our two possible two-level experiment configurations. We use the mixture phase for resynthesis.

\noindent\textbf{Loss functions}:
Among the possible spectrogram, waveform, and mask-based loss functions proposed in \cite{Petermann2023ICASSP_hyper}, we opt to use the cross-entropy (CE) loss to compare the ground-truth binary masks and the prediction as in classification setups \cite{WangDL2018ieeeacmaslp,Subramanian2019student}. In other words, we assume each TF bin embedding to belong to one source only and use a one-hot vector encoding (or IBM in the two-source mixture case) as the target mask on CE. \Cref{eq:loss1,eq:loss2,eq:loss3} denote the loss functions for three cases: the parent level, near sources and far sources in the children level, respectively: 
\begin{align}
    \text{Parents: }& \mathcal{L}^\text{parent}={\frac{1}{TF}}\sum_{t,f}\text{CE}\left(M^\text{parent}_{t,f,:}|| \hat{M}^\text{parent}_{t,f,:}\right) \label{eq:loss1}\\
    \text{Near children: }& \mathcal{L}^\text{near}={\frac{1}{TF}}\sum_{t,f}\text{CE}\left(M^\text{near}_{t,f,k\in\mathcal{N}}|| \hat{M}^\text{near}_{t,f,k\in\mathcal{N}}\right) \label{eq:loss2}\\
    \text{Far children: }& \mathcal{L}^\text{far}={\frac{1}{TF}}\sum_{t,f}\text{CE}\left(M^\text{far}_{t,f,k\in\mathcal{F}}|| \hat{M}^\text{far}_{t,f,k\in\mathcal{F}}\right) \label{eq:loss3},
\end{align}
where the parent-level loss effectively compares the source estimate to the sum of near sources group, e.g., $x^{(\mathcal{N})}\approx \text{iSTFT}(\hat{M}^\text{parent}_{:,:,0}\odot X)$, and ditto for the far sources. In contrast, the child-level loss functions are parent-specific, meaning the source separation results are compared to the child sources that belong to the same parent, either near or far. 
Note that while the order of the parent source groups is assumed, permutation invariant training (PIT) \cite{KolbaekM2017upit} for the child-level is required as each parents' child sources are in no particular order. Hence, the $\mathcal{L}^\text{parent}$ works as the sole loss function for parent-only source separation task as in \cite{Patterson2022DistSS}, i.e., the one-level model. Meanwhile, our two-level hierarchical model is trained from the compound of all loss terms: $\Loss = \Loss^\text{parent} + \Loss^\text{near} + \Loss^\text{far}$.

\noindent\textbf{Training}:
All experiments use the Adam optimizer for the Euclidean parameters, and the Riemannian Adam \cite{becigneul2018riemannian} implementation from \emph{geoopt} \cite{kochurov2020geoopt} for the hyperbolic parameters. All models are trained using chunks of 6.0 s, batch size of 96 over a total of 200 and 300 epochs for the one-level and two-level models, respectively. We use an initial learning rate of $10^{-3}$, which is halved if the validation loss does not improve for 10 epochs. We set the STFT size to 32-ms with a 50\% overlap and the square-root Hann window.

\noindent\textbf{Evaluation}:
Our main focus is to show the predictive uncertainty naturally conveyed by the hyperbolic space and how this information could potentially be exploited on downstream systems. 
We report the validity of the proposed method in terms of SI-SDR improvement (SI-SDRi) \cite{LeRouxJL2018sisdr} and use the best permutation at the child-level.
If one of the parents is silent (i.e., all speakers belong to either the ``near'' or ``far'' field), we evaluate the predicted silent sources by how much signal was bled into it. For that reason, we report the loudness ratio between the input mixture and the predicted silent source, called \textit{noise reduction}, as in \cite{Patterson2022DistSS} (i.e., the larger the better).

\begin{figure*}[thp]
    \centering
    \begin{subfigure}{0.325\textwidth}
    \includegraphics[width=1.0\linewidth]{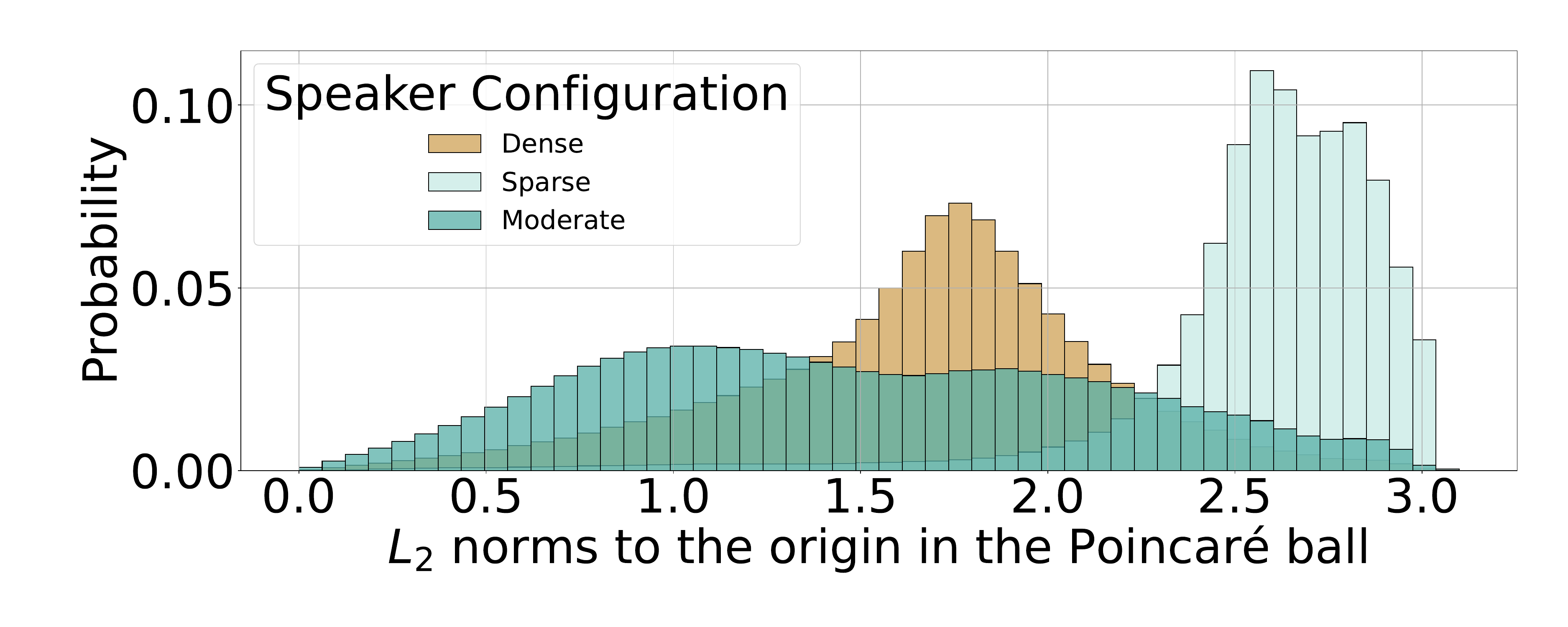}\vspace{-0.4cm}
    \caption{Each histogram denotes various speaker density configurations (covered in Table~\ref{table:dataset_stats}).}
    \label{fig:multiucs}
    \end{subfigure}
    \hfill
    \begin{subfigure}{0.325\textwidth}
    \includegraphics[width=1.0\linewidth]{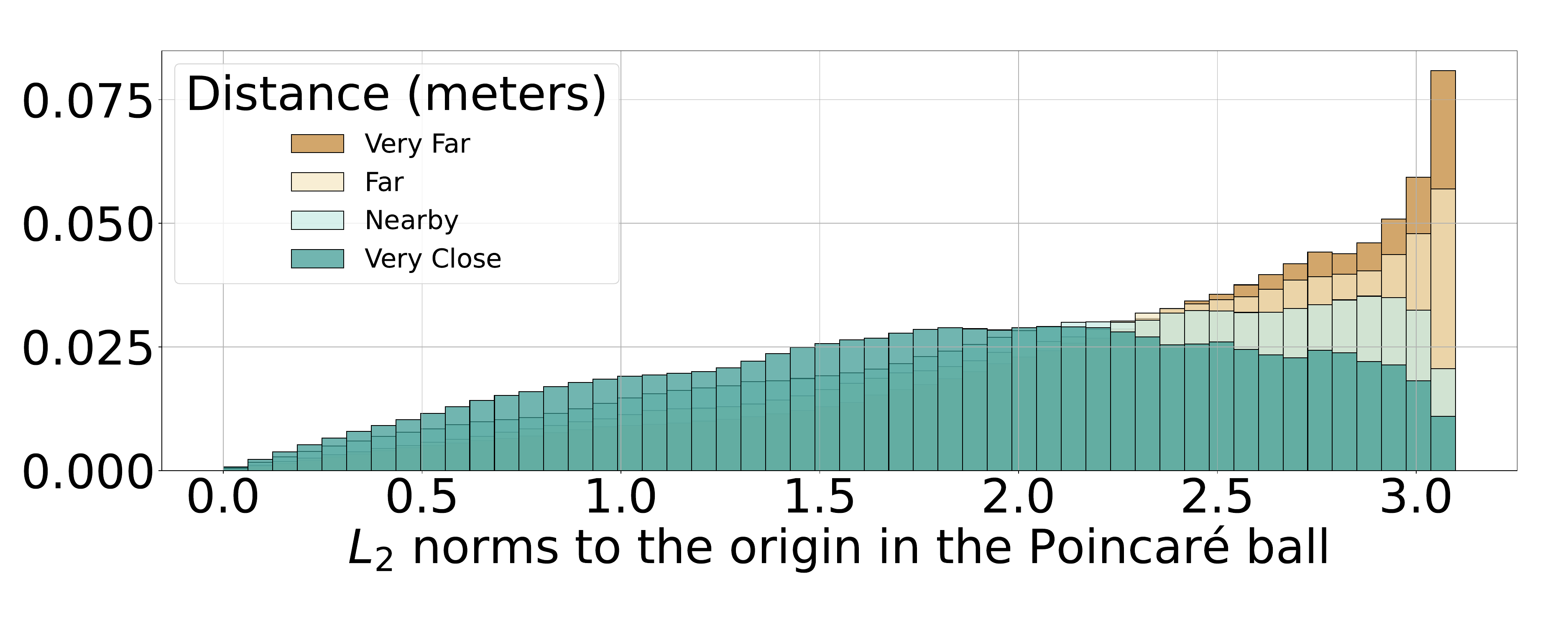}\vspace{-0.4cm}
    \caption{Each histogram denotes specific distance ranges between ``near'' and ``far'' sources.}
    \label{fig:near2far}
    \end{subfigure}
    \hfill
    \begin{subfigure}{0.325\textwidth}
    \includegraphics[width=1.0\linewidth]{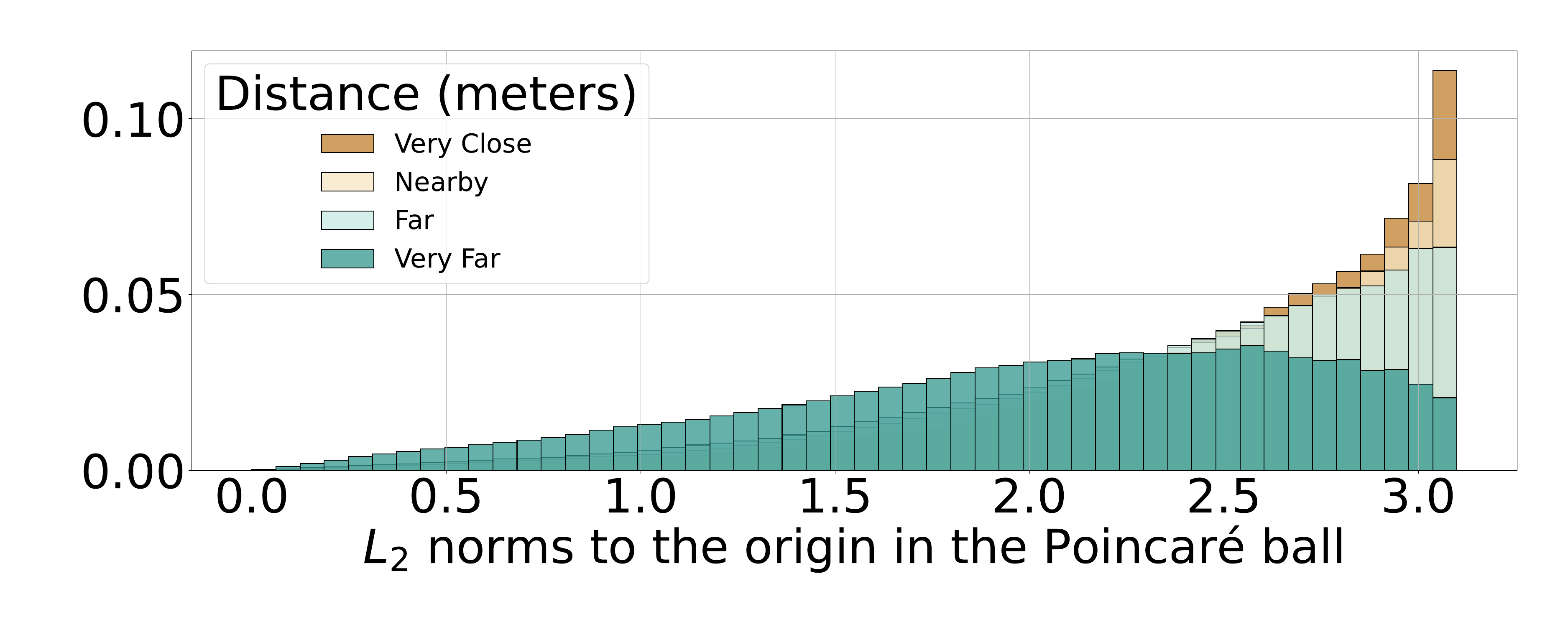}\vspace{-0.4cm}
    \caption{Each histogram denotes specific distance ranges between the microphone and nearest source.}
    \label{fig:near2mic}
    \end{subfigure}
    \captionsetup{belowskip=-.3cm}
    \vspace{-0.2cm}
    \caption{Distributions of $L_2$ norms to the origin of the Poincar\'e ball for all embeddings in their dedicated testing-set, as a function of various acoustics paradigms, such as speakers configuration (left), sources (center), and microphone (right) distances.}
\end{figure*}

\section{Experimental Results}
\subsection{Uncertainty Analysis}

Our experiments explore how well the Poincar\'e ball can model the task of hierarchical distance-based source separation. Precisely, we investigate to what extent the hyperbolic space can tie the intrinsic notion of certainty to the acoustic configurations, such as source-to-microphone distance, the relative distance among speakers, and speaker density, with a fixed distance threshold $\tau=0.8$ meter. 

\noindent\textbf{Speaker density}: 
We refer to speaker density as the relation between the number of speakers present in the parent classes ``near'' and ``far''. For example, our two-level hierarchy (see Table~\ref{table:dataset_stats}), can have a ``dense" setup with up to two child sources on each parent side ($\{2,2\}$), while its ``moderate'' ($\{2,1\}$, $\{1,2\}$), and ``sparse'' ($\{2,0\}$, $\{0,2\}$) setups are also considered. 
Fig ~\ref{fig:multiucs} shows the embeddings' $L_2$ norm distribution across all three configurations. We note that when all children belong to one parent, the model showcases high level of certainty, with most of the projected embeddings located far from the origin.

\noindent\textbf{Distance among sources}: 
To observe how the relative distance among the sources can affect the hyperbolic certainty in the speakers' embeddings, we design a small test set with one child per parent, i.e., $x^{(\mathcal{N})}\!=\!s^{(1)}$ and $x^{(\mathcal{F})}\!=\!s^{(2)}$. These sources are equidistanced from the threshold $\tau$, meaning $d(s^{(2)})\!-\!\tau\!=\!\tau\!-\!d(s^{(1)})$, where $d(\cdot)$ denotes the distance from the microphone. We then vary $d(s^{(1)})$ and $d(s^{(2)})$ such that the two sources' relative distance $d(s^{(2)})\!-\!d(s^{(1)})$ increases from 0, i.e., $d(s^{(2)})\!=\!d(s^{(1)})\!=\!\tau$, to 1.4 meters, i.e., $d(s^{(2)})\!=\!1.5$ and $d(s^{(1)})\!=\!0.1$. Fig. ~\ref{fig:near2far} shows the observed distribution of speakers embeddings' $L_2$ norms as a function of their relative distance. We note an overall decrease in certainty as their relative distance decreases, and vice versa.

\noindent\textbf{Distance from the microphone:} 
We also consider the sources' distance from the microphone by designing another one-child-per-parent test set with the far source away from the microphone with a fixed distance $d(s^{(2)})=2.9$, while the near source $s^{(1)}$ varies its distance to the microphone in the range of $[0.2, 0.8]$. From Fig ~\ref{fig:near2mic}, we observe that the closer $s^{(1)}$ is to the microphone, the higher the certainty is observed from the hyperbolic embeddings.

\subsection{Separation performances}

\begin{table}[t]
\scriptsize
\centering
  \sisetup{
    detect-weight, %
    mode=text, %
    round-mode=places,
    round-precision=1,
    table-format=2.1,
    table-number-alignment=center
    }

\begin{tabular}[t]{l*{3}{S}}
\toprule
 \multicolumn{1}{c}{$c$} & \multicolumn{2}{c}{SI-SDRi} & \multicolumn{1}{c}{Noise Reduction} \\
 \cmidrule(lr){1-1} \cmidrule(lr){2-3} \cmidrule(lr){4-4}
 & \multicolumn{1}{c}{Near} & \multicolumn{1}{c}{Far} & \multicolumn{1}{c}{Near} \\
\midrule
No Proc & 3.07 & -3.07 & \textemdash \\
\midrule

0.0 & \bfseries 6.47 & 9.14 & \bfseries 27.1 \\
-0.1 & 6.25 & 9.02 & \bfseries 27.05 \\
-1.0 & 6.29 & \bfseries 9.18 & 25.75 \\
\bottomrule
\end{tabular}
\captionsetup{belowskip=-.4cm}
\caption{SI-SDR improvement (dB) comparisons on models involving different curvatures (i.e, $c=0.0$ for Euclidean and $c<0$ for hyperbolic) on a parents-only training setup. Power reduction is reported when ``near'' is silent (the larger the better).}
\label{table:single}
\end{table}
\begin{table}[t]
\scriptsize
\centering
  \sisetup{
    detect-weight, %
    mode=text, %
    round-mode=places,
    round-precision=1,
    table-format=2.1,
    table-number-alignment=center
    }
\begin{subtable}[t]{1.\linewidth}
\centering
\begin{tabular}[t]{l*{6}{S}}
\toprule

& \multicolumn{3}{c}{Parents} & \multicolumn{3}{c}{Children} \\
 \cmidrule(lr){2-4} \cmidrule(lr){5-7} 
Speaker Density & \multicolumn{1}{c}{$c \ = \ 0.0$} & \multicolumn{1}{c}{$-0.1$} & \multicolumn{1}{c}{$-1.0$} & \multicolumn{1}{c}{$0.0$} & \multicolumn{1}{c}{$-0.1$} & \multicolumn{1}{c}{$-1.0$}\\

\midrule

2 near, 0 far & \bfseries 31.81$^\textbf{\dag}$ & 29.75$^\textbf{\dag}$ & 28.65$^\textbf{\dag}$ & 8.36 & 8.36 & \bfseries 8.42 \\
2 near, 1 far & 1.48 & 1.48 & 1.46 & 5.87 & 6.06 & \bfseries 6.28 \\
2 near, 2 far & \bfseries 2.36 & 2.31 & 2.28 & 3.96 & \bfseries 4.12 & 4.05 \\
1 near, 2 far & 6.66 & 6.72 & \bfseries 6.77 & 7.12 & \bfseries 7.52 & 7.05 \\
0 near, 2 far & \bfseries 63.26$^\textbf{\dag}$ & 61.33$^\textbf{\dag}$ & 63.17$^\textbf{\dag}$ & 8.91 & \bfseries 9.16 & \bfseries 9.15 \\
\midrule
Average (SI-SDRi) & 3.5 & 3.5 & 3.5 & 6.84 & \bfseries 7.04 & \bfseries 6.99 \\

\bottomrule

3 near, 0 far & 14.17$^\textbf{\dag}$ & 19.48$^\textbf{\dag}$ & 15.72$^\textbf{\dag}$ & 5.66 & 4.9 & \bfseries 6.03 \\
3 near, 1 far & 0.58 & \bfseries 0.79 & 0.72 & 4.19 & 4.02 & \bfseries 4.4 \\
2 near, 2 far & 2.48 & 2.51 & \bfseries 2.62 & 4.61 & 4.61 & \bfseries 4.91 \\
1 near, 3 far & 4.87 & 4.66 & \bfseries 4.96 & 5.02 & 5.07 & \bfseries 5.43 \\
0 near, 3 far & \bfseries 49.21$^\textbf{\dag}$ & 42.56$^\textbf{\dag}$ & 41.62$^\textbf{\dag}$ & 5.93 & 6.93 & \bfseries 6.98 \\
\midrule
Average (SI-SDRi) & 2.64 & 2.65 & \bfseries 2.76 & 5.08 & 5.1 & \bfseries 5.55 \\

\bottomrule
\end{tabular}
\end{subtable}.\vspace{-.2cm}
\captionsetup{belowskip=-.4cm}
\caption{SI-SDR improvement (dB) of the models involving different curvatures (i.e, $c=0.0$ for Euclidean and $c<0.0$ for hyperbolic) on our 2-level setup with up to two and three child sources per parent (top and bottom rows, respectively). Noise Reduction, when one of the parent is silent, is indicated by $\textbf{\dag}$ (the larger the better).}
\label{table:multi_22}
\end{table}\vspace{-.0cm}

Table~\ref{table:single} reports the results in terms of SI-SDR improvement from our one-level model (i.e., the parent-only case), where no performance gain is induced by the hyperbolic setup. We speculate that it is due to the lack of hierarchy in the problem setup, negating the benefit from the hyperbolic manifold. However, we emphasize that as shown in Fig. \ref{fig:main}, the notion of certainty in the proposed hyperbolic approach can provide an additional user interface, identifying borderline cases. 

Table~\ref{table:multi_22} shows the results from the two-level setups. We report the performance of both Euclidean ($c=0.0$) and hyperbolic models ($c \in [-0.1,-1.0]$). In the top rows (two children sources per parent), the parental separation results are not affected by our hyperbolic setup. However, we observe clearer improvement in the child-level results using the hyperbolic method, with an average improvement of 0.2 dB SI-SDR over their Euclidean counterpart. The more complex two-level setup with up to three child sources per parent (bottom rows) shows a solid 0.5 dB SI-SDR improvement with $c=-1.0$. We believe that the children benefit the most from the hyperbolic treatment due to the intrinsic nature of the space to efficiently models the hierarchical structure modelled by the data. This could also explain the greater gain when more children are present. 

This relatively small performance margin from the hyperbolic method aligns well with the literature \cite{Atigh_2022_CVPR,Khrulkov_2020_CVPR,Petermann2023ICASSP_hyper}. In this work, we rather bring the readers' attention to the explanatory and intuitive nature of the hyperbolic representation. For example, obtaining a natural measure of uncertainty predictions for these hierarchical tasks could help tackle the typical trade-off between sensitivity and specificity: one can still choose to acquire the closer source group if the system is uncertain about the child-level source separation results. 

\section{Conclusions}
\label{sec:conclusion}

In this work, we redefined distance-based single-channel speech separation as a hierarchical task performed on an hyperbolic manifold (i.e., the Poincar\'e ball). We designed a two-level hierarchy, whose parent level distinguishes sources based on their distance (i.e., near and far) to the microphone, while each parent may contain up to two or three speakers (i.e., children). We empirically showed that the hyperbolic space consistently outperformed our Euclidean baseline when the task was challenging enough with more child-level sources. In addition, the network was successful in modeling some key acoustic concepts such as speaker density, the relative distance between sources, or microphones, and translating them to the notion of hyperbolic certainty which could be effectively used towards practical downstream tasks. Code will be made available~\footnote{\url{https://minjekim.com/research-projects/hdss}}.

\bibliographystyle{IEEEtran}
\bibliography{mjkim}

\begin{thebibliography}{10}
\providecommand{\url}[1]{#1}
\csname url@samestyle\endcsname
\providecommand{\newblock}{\relax}
\providecommand{\bibinfo}[2]{#2}
\providecommand{\BIBentrySTDinterwordspacing}{\spaceskip=0pt\relax}
\providecommand{\BIBentryALTinterwordstretchfactor}{4}
\providecommand{\BIBentryALTinterwordspacing}{\spaceskip=\fontdimen2\font plus
\BIBentryALTinterwordstretchfactor\fontdimen3\font minus
  \fontdimen4\font\relax}
\providecommand{\BIBforeignlanguage}[2]{{%
\expandafter\ifx\csname l@#1\endcsname\relax
\typeout{** WARNING: IEEEtran.bst: No hyphenation pattern has been}%
\typeout{** loaded for the language `#1'. Using the pattern for}%
\typeout{** the default language instead.}%
\else
\language=\csname l@#1\endcsname
\fi
#2}}
\providecommand{\BIBdecl}{\relax}
\BIBdecl

\bibitem{mcdermott2009cpp}
J.~H. McDermott, ``The cocktail party problem,'' \emph{Current Biology},
  vol.~19, no.~22, pp. R1024--R1027, 2009.

\bibitem{Karns2015AuditoryAI}
C.~M. Karns, E.~Isbell, R.~J. Giuliano, and H.~J. Neville, ``Auditory attention
  in childhood and adolescence: An event-related potential study of spatial
  selective attention to one of two simultaneous stories,'' \emph{Developmental
  Cognitive Neuroscience}, vol.~13, pp. 53 -- 67, 2015.

\bibitem{WangDL2018ieeeacmaslp}
D.~L. {Wang} and J.~{Chen}, ``{Supervised Speech Separation Based on Deep
  Learning: An Overview},'' \emph{IEEE/ACM Transactions on Audio, Speech, and
  Language Processing}, vol.~26, no.~10, pp. 1702--1726, 2018.

\bibitem{liu2022separate}
X.~Liu \emph{et~al.}, ``{Separate What You Describe: Language-Queried Audio
  Source Separation},'' in \emph{Proc. Interspeech}, 2022.

\bibitem{tzinis2022heterogeneous}
E.~Tzinis \emph{et~al.}, ``{Heterogeneous Target Speech Separation},'' in
  \emph{Proc. Interspeech}, 2022, pp. 1796--1800.

\bibitem{darwin2000selective}
C.~J. Darwin and R.~W. Hukin, ``Effectiveness of spatial cues, prosody, and
  talker characteristics in selective attention,'' \emph{The Journal of the
  Acoustical Society of America}, vol. 107, no.~2, pp. 970--977, 2000.

\bibitem{rongzhi2019spatial}
G.~Rongzhi \emph{et~al.}, ``Neural spatial filter: Target speaker speech
  separation assisted with directional information,'' in \emph{Proc.
  Interspeech}, 09 2019, pp. 4290--4294.

\bibitem{heitkaemper2019position}
J.~Heitkaemper, T.~Feh{\'e}r, M.~Freitag, and R.~Haeb-Umbach, ``A study on
  online source extraction in the presence of changing speaker positions,'' in
  \emph{Statistical Language and Speech Processing}, C.~Mart{\'i}n-Vide,
  M.~Purver, and S.~Pollak, Eds.\hskip 1em plus 0.5em minus 0.4em\relax
  Springer International Publishing, 2019, pp. 198--209.

\bibitem{EskimezSE2021personalized}
S.~E. Eskimez \emph{et~al.}, ``{Personalized Speech Enhancement: New Models and
  Comprehensive Evaluation},'' in \emph{Proc. of the IEEE International
  Conference on Acoustics, Speech, and Signal Processing (ICASSP)}, 2022, pp.
  356--360.

\bibitem{Patterson2022DistSS}
K.~Patterson, K.~W. Wilson, S.~Wisdom, and J.~R. Hershey, ``Distance-based
  sound separation,'' in \emph{Interspeech 2022, 23rd Annual Conference of the
  International Speech Communication Association, Incheon, Korea, 18-22
  September 2022}, H.~Ko and J.~H.~L. Hansen, Eds.\hskip 1em plus 0.5em minus
  0.4em\relax {ISCA}, 2022, pp. 901--905.

\bibitem{Delcroix2018speaker_beam}
M.~Delcroix \emph{et~al.}, ``{Single Channel Target Speaker Extraction and
  Recognition with Speaker Beam},'' in \emph{Proc. of the IEEE International
  Conference on Acoustics, Speech, and Signal Processing (ICASSP)}, 2018, pp.
  5554--5558.

\bibitem{Atigh_2022_CVPR}
M.~G. Atigh \emph{et~al.}, ``Hyperbolic image segmentation,'' in \emph{Proc. of
  the IEEE International Conference on Computer Vision and Pattern Recognition
  (CVPR)}, Jun. 2022.

\bibitem{Khrulkov_2020_CVPR}
V.~Khrulkov \emph{et~al.}, ``Hyperbolic image embeddings,'' in \emph{Proc. of
  the IEEE International Conference on Computer Vision and Pattern Recognition
  (CVPR)}, June 2020.

\bibitem{shimizu2021hyperbolic}
R.~Shimizu, Y.~Mukuta, and T.~Harada, ``Hyperbolic neural networks++,'' in
  \emph{International Conference on Learning Representations}, 2021.

\bibitem{Petermann2023ICASSP_hyper}
D.~Petermann, G.~Wichern, A.~Subramanian, and J.~{Le Roux}, ``Hyperbolic audio
  source separation,'' in \emph{Proc. of the IEEE International Conference on
  Acoustics, Speech, and Signal Processing (ICASSP)}, Jun. 2023.

\bibitem{Nakashima2022VAE}
F.~Nakashima \emph{et~al.}, ``Hyperbolic timbre embedding for musical
  instrument sound synthesis based on variational autoencoders,'' in
  \emph{Proc. APSIPA ASC}, 2022.

\bibitem{GermainF2023hyperbolic}
F.~Germain, G.~Wichern, and J.~{Le Roux}, ``Hyperbolic unsupervised anomalous
  sound detection,'' in \emph{Proc. of the IEEE Workshop on Applications of
  Signal Processing to Audio and Acoustics (WASPAA)}, 2023.

\bibitem{sarkar2011distortion}
R.~Sarkar, ``Low distortion delaunay embedding of trees in hyperbolic plane,''
  in \emph{Proc. International Symposium On Graph Drawing}, 2012, pp. 355--366.

\bibitem{suris2021hyperfuture}
D.~Sur\'is, R.~Liu, and C.~Vondrick, ``Learning the predictability of the
  future,'' in \emph{Proc. of the IEEE International Conference on Computer
  Vision and Pattern Recognition (CVPR)}, 2021.

\bibitem{LeRouxJL2018sisdr}
J.~{Le Roux}, S.~Wisdom, H.~Erdogan, and J.~R. Hershey, ``{SDR --} half-baked
  or well done?'' in \emph{Proc. of the IEEE International Conference on
  Acoustics, Speech, and Signal Processing (ICASSP)}, 2019.

\bibitem{fang2023uncertainty}
H.~Fang, D.~Becker, S.~Wermter, and T.~Gerkmann, ``Integrating uncertainty into
  neural network-based speech enhancement,'' \emph{IEEE/ACM Transactions on
  Audio, Speech, and Language Processing}, vol.~31, pp. 1587--1600, 2023.

\bibitem{nickel2017poincare}
M.~Nickel and D.~Kiela, ``Poincar\'{e} embeddings for learning hierarchical
  representations,'' in \emph{Advances in Neural Information Processing
  Systems}, I.~Guyon \emph{et~al.}, Eds., vol.~30.\hskip 1em plus 0.5em minus
  0.4em\relax Curran Associates, Inc., 2017.

\bibitem{chen2022fully}
W.~Chen \emph{et~al.}, ``Fully hyperbolic neural networks,'' in
  \emph{Proceedings of the 60th Annual Meeting of the Association for
  Computational Linguistics (Volume 1: Long Papers)}.\hskip 1em plus 0.5em
  minus 0.4em\relax Dublin, Ireland: Association for Computational Linguistics,
  May 2022, pp. 5672--5686.

\bibitem{scheibler2018pyroomacoustics}
R.~Scheibler, E.~Bezzam, and I.~Dokmani\'{c}, ``Pyroomacoustics: A python
  package for audio room simulation and array processing algorithms,'' in
  \emph{Proc. of the IEEE International Conference on Acoustics, Speech, and
  Signal Processing (ICASSP)}, 2018, p. 351–355.

\bibitem{librilight}
J.~{Kahn} \emph{et~al.}, ``Libri-light: A benchmark for asr with limited or no
  supervision,'' in \emph{ICASSP 2020 - 2020 IEEE International Conference on
  Acoustics, Speech and Signal Processing (ICASSP)}, 2020, pp. 7669--7673.

\bibitem{PanayotovV2015Librispeech}
V.~Panayotov, G.~Chen, D.~Povey, and S.~Khudanpur, ``{Librispeech}: {An} {ASR}
  corpus based on public domain audio books,'' in \emph{Proc. of the IEEE
  International Conference on Acoustics, Speech, and Signal Processing
  (ICASSP)}, 2015, pp. 5206--5210.

\bibitem{Subramanian2019student}
A.~Subramanian, S.~J. Chen, and S.~Watanabe, ``Student-teacher learning for
  blstm mask-based speech enhancement,'' in \emph{Proc. Interspeech}, 09 2018,
  pp. 3249--3253.

\bibitem{KolbaekM2017upit}
M.~Kolb{\ae}k, D.~Yu, Z.~H. Tan, and J.~Jensen, ``Multitalker speech separation
  with utterance-level permutation invariant training of deep recurrent neural
  networks,'' \emph{IEEE/ACM Transactions on Audio, Speech, and Language
  Processing}, vol.~25, no.~10, pp. 1901--1913, 2017.

\bibitem{becigneul2018riemannian}
H.~Kasai, P.~Jawanpuria, and B.~Mishra, ``{R}iemannian adaptive stochastic
  gradient algorithms on matrix manifolds,'' in \emph{Proc. of the
  International Conference on Machine Learning (ICML)}, Jun. 2019.

\bibitem{kochurov2020geoopt}
M.~Kochurov, R.~Karimov, and S.~Kozlukov, ``Geoopt: {R}iemannian optimization
  in {PyTorch},'' \emph{arXiv preprint arXiv:2005.02819}, 2020.

\end{thebibliography}

\end{document}